\documentclass{article}

\usepackage{arxiv}

\usepackage[utf8]{inputenc} % allow utf-8 input
\usepackage[T1]{fontenc}    % use 8-bit T1 fonts
\usepackage{hyperref}       % hyperlinks
\usepackage{url}            % simple URL typesetting
\usepackage{booktabs}       % professional-quality tables
\usepackage{amsfonts}       % blackboard math symbols
\usepackage{nicefrac}       % compact symbols for 1/2, etc.
\usepackage{microtype}      % microtypography
\usepackage{lipsum}		% Can be removed after putting your text content
\usepackage{graphicx}
\def\BibTeX{{\rm B\kern-.05em{\sc i\kern-.025em b}\kern-.08em
    T\kern-.1667em\lower.7ex\hbox{E}\kern-.125emX}}
\usepackage{doi}
\usepackage{tabularx}
\usepackage{acronym}

\acrodef{iot}[IoT]{Internet of Things}
\acrodef{iiot}[IIoT]{Industrial Internet of Things}
\acrodef{mno}[MNO]{Mobile Network Operators}
\acrodef{urllc}[uRLLC]{Ultra Reliable and Low Latency Communication}
\acrodef{xurllc}[xURLLC]{extreme Ultra Reliable and Low Latency Communication}
\acrodef{cpps}[CPPS]{cyber-physical production systems}
\acrodef{qos}[QoS]{Quality of Service}
\acrodef{nm}[NetEm]{network emulation}
\acrodef{nin}[NiN]{Network-in-Network}
\acrodef{ncs}[NCS]{Networked Control System}
\acrodef{embb}[eMBB]{Enhanced Mobile Broadband}
\acrodef{kpis}[KPIs]{Key Performance Indicators}
\acrodef{dsm}[DSM]{Dynamic Spectrum Management}
\acrodef{cnc}[CNC]{Computerized Numerical Control}
\acrodef{fpga}[FPGA]{Field Programmable Gate Array}
\acrodef{cpf}[CPF]{control plane fabric}
\acrodef{dht}[DHT]{distributed hash table}
\acrodef{npn}[NPN]{Non-Public Network}
\acrodef{ai}[AI]{Artificial Intelligence}
\acrodef{rpi}[RPI]{Raspberry Pi}
\acrodef{sm}[SM]{spectrum manager}
\acrodef{sn}[SN]{sub-network}
\acrodef{snc}[SNC]{sub-network controller}
\acrodef{dt}[DT]{Digital Twin}
\acrodef{agv}[AGV]{Automated Guided Vehicle}
\acrodef{pn}[PN]{private network}
\acrodef{nin}[NiN]{Networks-in-Network}
\acrodef{mrat}[Multi-RAT]{multiple radio access technology}
\title{Adaptive 6G Networks-in-Network Management for Industrial Applications}

\author{ {Daniel~Lindenschmitt}\\
	Institute for Wireless Communication \\and Navigation\\
	RPTU Kaiserslautern-Landau\\
	\texttt{daniel.lindenschmitt@rptu.de} \\
	\And
	{Paul~Seehofer} \\
	Institute of Telematics\\
	Karlsruhe Institute of Technology\\
	\texttt{paul.seehofer@kit.edu} \\
    \And
	{Marius~Schmitz} \\
	Institute for Manufacturing Technology \\and Production Systems\\
	RPTU Kaiserslautern-Landau\\
	\texttt{marius.schmitz@rptu.de} \\
     \And
	{Jan~Mertes} \\
	Institute for Manufacturing Technology \\and Production Systems\\
	RPTU Kaiserslautern-Landau\\
	\texttt{jan.mertes@rptu.de} \\
 	\And
	{Roland~Bless} \\
	Institute of Telematics\\
	Karlsruhe Institute of Technology\\
	\texttt{roland.bless@kit.edu} \\
    \And
	{Martina~Zitterbart} \\
	Institute of Telematics\\
	Karlsruhe Institute of Technology\\
	\texttt{zitterbart@kit.edu} \\
    \And
	{Jan C.~Aurich} \\
	Institute for Manufacturing Technology \\and Production Systems\\
	RPTU Kaiserslautern-Landau\\
	\texttt{jan.aurich@rptu.de} \\
 \And
	{Hans D.~Schotten}\\
	Institute for Wireless Communication \\and Navigation\\
	RPTU Kaiserslautern-Landau\\
	\texttt{schotten@rptu.de} \
}

\date{}

\renewcommand{\shorttitle}{\textit{arXiv} Template}

\begin{document}
\maketitle

\begin{abstract}
This paper presents the application of \ac{dsm} for future 6G industrial networks, establishing an efficient controller for the \ac{nin} concept. The proposed architecture integrates nomadic as well as static \acp{sn} with diverse \ac{qos} requirements within the coverage area of an overlayer network, managed by a centralized \ac{sm}. Control plane connectivity between the \acp{sn} and the \ac{dsm} is ensured by the self-organizing KIRA routing protocol. The demonstrated system enables scalable, zero-touch connectivity and supports nomadic \acp{sn} through seamless discovery and reconfiguration. \acp{sn} are implemented for modular \ac{iiot} scenarios, as well as for mission-critical control loops and for logistics or nomadic behavior. The \ac{dsm} framework dynamically adapts spectrum allocation to meet real-time demands while ensuring reliable operation. The demonstration highlights the potential of \ac{dsm} and \acp{nin} to support flexible, dense, and heterogeneous wireless deployments in reconfigurable manufacturing environments.
\end{abstract}

\keywords{6G, Networks-in-Network, Spectrum-Sharing, Multi-RAT}

\section{Introduction}
The transition from the current 5G standard to the next-generation 6G standard will once again enable a significant increase in performance, such as data rates, and, more importantly, the integration of new functionalities to the mobile network. However, the available spectrum for data transmission remains the limiting factor for these \ac{kpis}. This limitation can be addressed either by allocating new frequency bands or by increasing spectral efficiency in the coverage area. In particular, this severely restricts the potential applications of cellular communication technologies in the context of \acp{npn}.

The technology presented in this work is based on previous work~\cite{10741442} and was significantly extended by the integration of an adaptive behavior of the \ac{dsm} for nomadic \acp{sn}. It integrates prior research on \acp{nin}~\cite{sublayer} and control plane connectivity for 6G~\cite{kira} into a unified \ac{dsm} approach. By using so-called \acp{nin}, \acp{sn} can be created for specific applications, such as industrial use cases. The demonstrations performs an ad-hoc reconfiguration of static and nomadic \acp{sn} available in the coverage area via a dynamic \ac{sm} system. The \acp{sn} have significantly smaller coverage areas and lower bandwidths compared to regular 5G \ac{npn}. Depending on their \ac{qos} requirements, this will allow a large number of \acp{sn} to coexist within a shared coverage area, called overlayer network. The use of mobile and nomadic \acp{sn}, as implemented in this demonstration, further increases the demands on an efficient \ac{snc}.  An \ac{agv} is virtually called to a machine tool via a call button, e.g. to deliver or collect a workpiece. \ac{dsm} recognizes the arriving \ac{agv} and efficiently distributes the available spectrum for all \acp{sn}, thus ensuring compliance with the \ac{qos} requirements.

\section{Background}
\label{background}
\subsection{Industrial Application in Network-in-Networks}
A wide range of wireless solutions is currently used for data transmission, particularly in \ac{iiot}, each offering specific advantages and disadvantages. Since 5G standardization, data transmission via \acp{npn} has also been one of these options. Here, 5G offers advantages in terms of latency and packet loss. With the evolution toward 6G, the number of supported use cases is expected to increase significantly \cite{6G_pers}. Research on the \ac{nin} concept has demonstrated that \ac{xurllc} requirements can be met effectively using \acp{nin}~\cite{sublayer}. Consequently, \acp{sn} are expected to play a key role in the development of organic, nomadic, and ad-hoc 6G networks~\cite{Organic6G, Scaling, 10849687}. This direction is further supported by recent work on \ac{mrat} applications, particularly in the context of \acp{npn}~\cite{10903794}, where end devices can access various protocols through a central interface. The demonstration presented here implements an alternative approach in which the \ac{sm} modifies the frequency allocation of the \acp{sn}, enabling data exchange via a token-based Wi-Fi protocol.

\subsection{Connectivity and Discovery using KIRA}
Distributed control plane applications such as \ac{dsm} require control plane connectivity for coordination between their components~\cite{rfc8368}.
The routing architecture KIRA~\cite{kira} is especially designed to provide such connectivity in large-scale and dynamic infrastructures in a resilient fashion: (1) it is scalable and zero-touch, e.g., does not require any configuration, allowing for a resilient and self-organized operation of the \ac{dsm} and (2) it supports node mobility and \emph{nomadic networks}, e.g., moving network partitions, out of the box, ensuring connectivity between all components even in highly dynamic environments with, e.g., moving \acp{sn}.
Furthermore, KIRA also includes a distributed hash table that can be used as a service discovery mechanism to enable the \acp{sn} to discover the central \ac{sm}.

\section{Application Scenario: Industrial IoT}
\label{usecase}

The modern manufacturing systems require the fusion of physical manufacturing processes with intelligent functions. A key enabler for this paradigm is the \ac{iiot} that enables the merge of the digital and physical world \cite{Boyes2018}.
This paradigm shift moves manufacturing away from static production lines towards highly flexible, modular environments capable of cost-efficiently producing customized goods \cite{ElMaraghy2021}. Such adaptivity is dependent on the deployment of wireless devices -- such as sensors, actuators, and controllers for mobile machines -- that must coexist and communicate reliably. This densification of wireless communication within a shop floor causes a complex and congested spectrum environment, posing a significant challenge to the operational functionality of the overall manufacturing system \cite{Vitturi2019}. The demonstration, therefore, addresses this problem by showcasing an intelligent, adaptive \ac{sm} framework for \ac{iiot} and machine control designed to support a dynamic and reconfigurable manufacturing system.

\begin{figure}[b]
    \centering
    \includegraphics[width=1.0\linewidth]{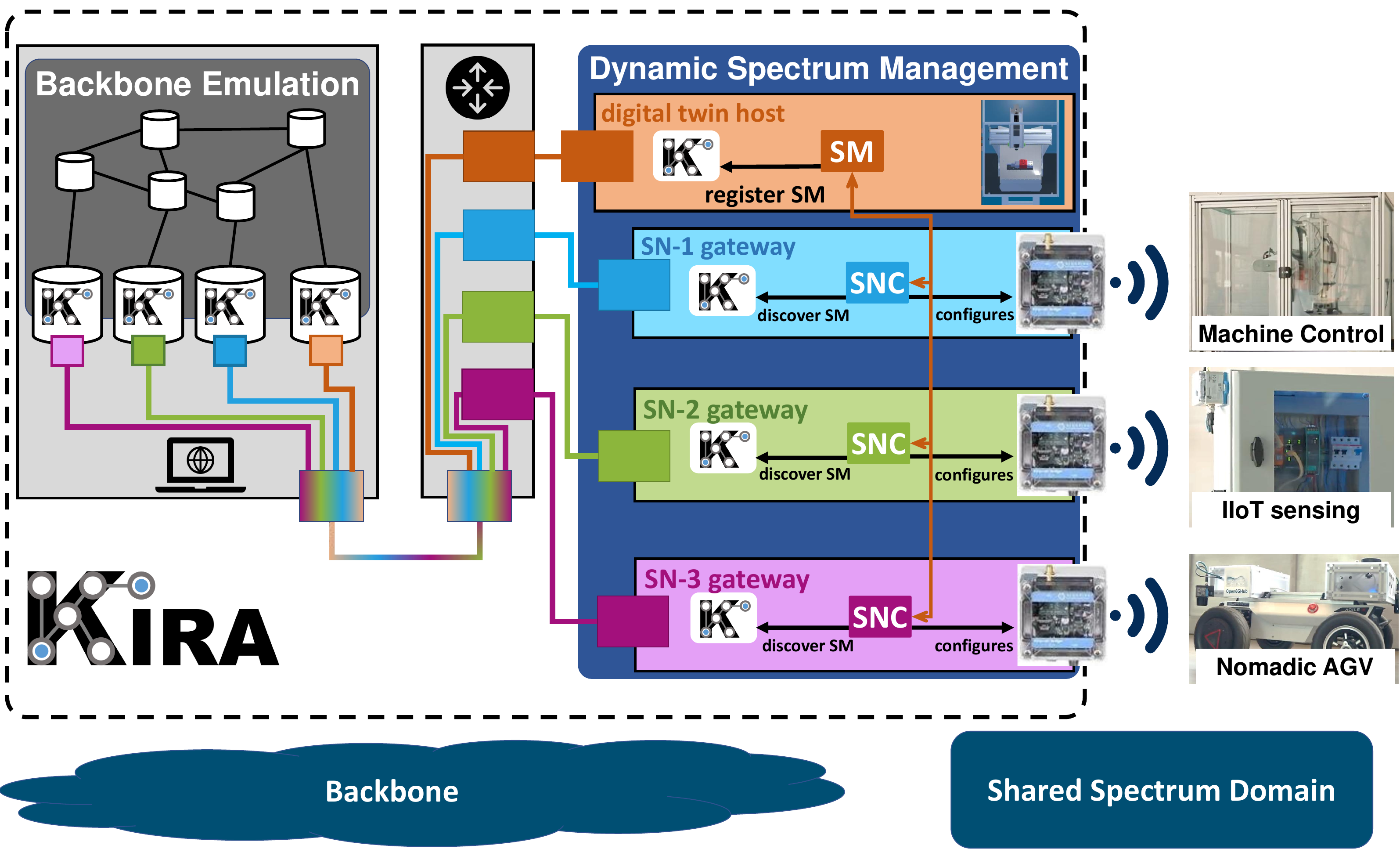}
    \caption{System architecture illustrating \acp{sn} and KIRA control plane}
    \label{fig:demo-architecture}
\end{figure}

The application scenario models a modular \ac{iiot} environment in manufacturing (interference caused by industrial equipment is not considered). It consists of one 5G \ac{npn} for holistic interconnectivity and three functionally distinct \acp{sn}, as shown in Figure \ref{fig:demo-architecture}, whose different performance requirements and operational patterns require dynamic spectrum management:

\begin{itemize}
    \item \emph{\ac{sn}-1 for mission-critical control systems:} This \ac{sn} handles traffic of the field device layer (actuators \& controllers) of the \ac{iiot} system. In this demonstration it connects the machine tool to its virtualized \ac{cnc}. The integrity of this link is mandatory for manufacturing quality and safety, demanding deterministic \ac{xurllc}.

    \item \emph{\ac{sn}-2 for high-fidelity \ac{iiot} sensing:} This \ac{sn} manages traffic for the sensors of the \ac{iiot} architecture, connecting various high-bandwidth sensors that monitor the manufacturing process. The \ac{iiot} data can then be used for central applications, such as the execution of a \ac{dt} of the production system. The interconnected sensors enable advanced analytics and predictive functions, e.g. for maintenance.

    \item  \emph{\ac{sn}-3 for nomadic logistics and material handling:} A key element of a flexible factory is automated logistics, represented here by an \ac{agv}. This nomadic \ac{sn} introduces a high degree of spatial and temporal uncertainty into the spectrum demand profile as the \ac{agv} moves through the manufacturing system. It requires robust connectivity for safe navigation while periodically demanding high throughput for task-related data exchange.
\end{itemize}
A conventional, static spectrum allocation is inadequate for this environment. The transient nature of the \ac{agv} and the a-periodic high-data-rate traffic from the sensor network require dynamic spectrum allocation depending on the currently coexisting \acp{sn}. 
\begin{figure}
    \centering
    \includegraphics[width=1.0\linewidth]{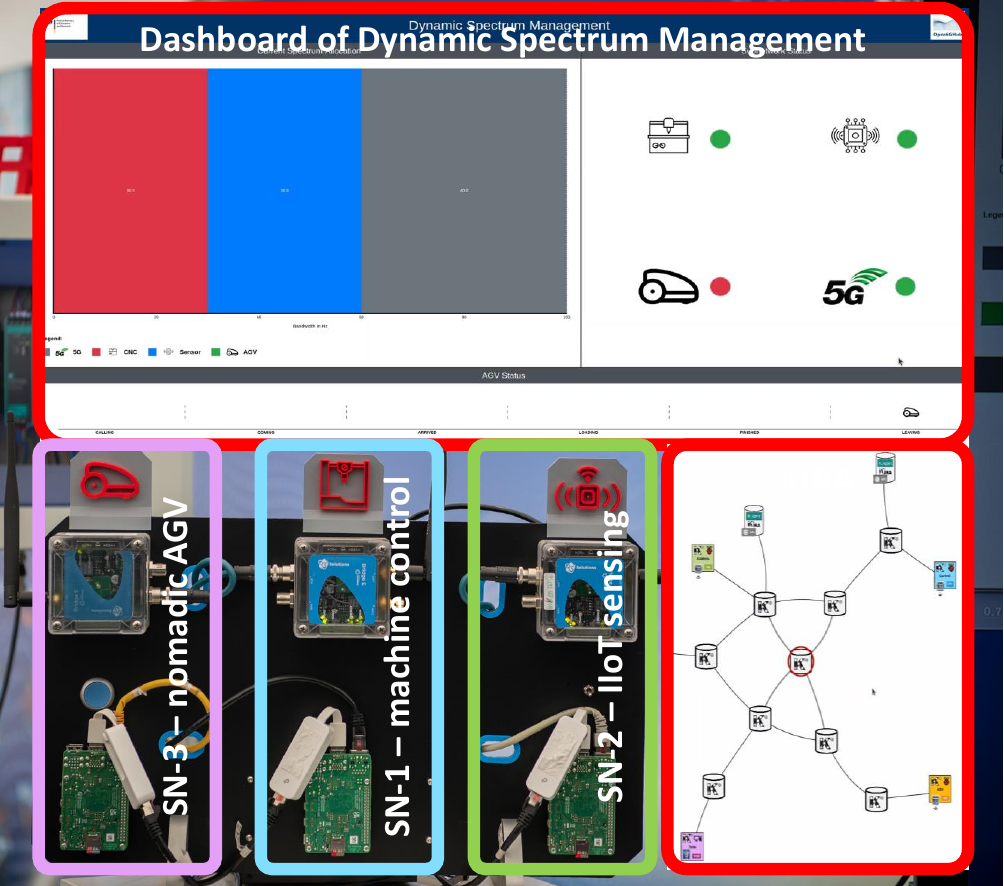}
    \caption{Overview of demonstration setup}
    \label{fig:demo-setup}
\end{figure}

\section{Demo Setup}
\label{demo}
The demonstrator of an ad-hoc \ac{nin} structure consisting of the following components, as shown in Figure~\ref{fig:demo-setup}:  \emph{1) \ac{sn} for closed loop control, 2) \ac{sn} for \ac{iiot} applications} and \emph{3) a \ac{sn} for a nomadic \ac{agv}.} All \acp{sn} are connected to the \ac{sm} via their \ac{snc}, which manages the connected \acp{sn}.
\begin{itemize}
    \item \emph{Zero-touch connectivity:} A KIRA routing daemon running on each node in the backbone network (including the \ac{sn} master nodes) ensures connectivity for this dynamic topology without any configuration. The \acp{snc} use KIRA's discovery mechanism, to discover the \ac{sm}, and its resilient connectivity to communicate, even when network conditions and participants change dynamically.

    \item \emph{Ad-hoc integration of nomadic \acp{sn}:} The \ac{dsm} orchestrates spectrum resources to support planned reconfigurations and respond to unplanned events without compromising performance. For instance, upon receiving intent data from the \ac{snc} that the \ac{agv} is scheduled to enter the cell, the \ac{sm} can proactively allocate spectrum to guarantee its seamless operation, potentially by temporarily downscaling the allocation for the less critical sensor network (SN-2) while preserving the mission-critical CNC control (SN-1). For nomadic \acp{sn} (SN-3) joining the overlayer network for the first time in an unplanned fashion, KIRA provides connectivity immediately, so that the \ac{snc} can discover, connect to and register with the \ac{sm} in order to get a frequency allocation. Moreover, KIRA's mobility support ensures connectivity for \acp{snc} of nomadic \acp{sn} that move through the backbone networks topology.

    \item \emph{\ac{mrat}-enabled communication:} The application of the \ac{dsm} approach enables support for Wi-Fi or other non-3GPP compliant radio technologies, if operated within the monitored frequency range. For this demonstration, the \acp{nin} are implemented using \acp{rpi}, with each \ac{rpi} serving as the master node for its associated \ac{sn}. A \ac{snc} running on each \ac{rpi} communicates with the \ac{sm} via the connectivity provided by KIRA, registers the frequency requirements, and receives an allocation response following a negotiation process. The custom-developed hardware is designed to operate with a token-based Wi-Fi protocol but transmits in the 3.7~GHz to 3.8~GHz range, outside the ISM bands.
\end{itemize}

%\begin{table}[h]
%\centering
%\caption{Dynamic spectrum allocation for \ac{nin}}
%    \begin{tabularx}{\linewidth}{lcc}
%        \hline        
%         &   \textbf{CNC Spectrum} & \textbf{Sensor Spectrum}  \\
%        \hline
%        Both running &  40\,MHZ & 60\,MHz \\
%        
%        Sensor Network unavailable & 100\,MHz & 0\,MHz \\ 
%        
%        CNC Network unavailable &  0\,MHz & 100\,MHz \\
%        
%        Both unavailable &  0\,MHz & 0\,MHz \\
%    \hline
%    \end{tabularx}
%    \label{tab:spectrum}
%\end{table}

\section{Demo Walkthrough}
\label{walk}
The sub-network SN-1 is responsible for communication between an industrial machine and the corresponding closed loop control, SN-2 transmits data between the \ac{iiot} sensors and the \ac{dt} and SN-3 is responsible for the nomadic \ac{agv}, which appears sporadically in the coverage area. The demonstration shows the ad-hoc reconfiguration of the \acp{sn} managed by the \ac{sm} and also integrates dynamic changes caused by new \acp{sn} such as the nomadic \ac{agv}. The presentation is divided into the following parts:
\begin{itemize}
    \item \emph{Reconfiguration:} Participants can interact with the system in two ways. The non-safety-relevant SN-2 data can be switched on or off, in addition the AGV (SN-3) can be virtually sent to the machine via a call button. The \ac{sm} triggers the network reconfiguration process and assigns new frequencies, which are then applied by the \ac{snc}.
    \item \emph{Visualization:} A dashboard in the backbone of the systems displays the current status of KIRA and all \acp{sn} attached to the \ac{dsm}. In addition, the current execution status of the nomadic \ac{agv} is shown as well as the \ac{dt} showing the \ac{cnc} machine with sensor data evaluation.
\end{itemize}
%Additional material of the demonstration is available here: \emph{nextcloud.eit.uni-kl.de/s/Kk8QENpFzaPioAx}

\section{Conclusion and Future Work}
\label{concl}
In this paper, the \ac{dsm} approach based on previous work~\cite{10741442} was further developed. By integrating an adaptive behavior of the \ac{dsm} for nomadic \acp{sn}, a solution for ad-hoc reconfiguration was developed for both static and nomadic \acp{sn} within the coverage area of an overlayer network. By integrating nomadic and static \acp{sn} into a shared spectrum domain of an overlayer network, the system addresses key challenges of dense and heterogeneous wireless deployments, opening up new potential applications for industrial environments using \acp{nin} and the future 6G standard. The combination of a centralized \ac{sm} with the self-organizing KIRA routing architecture enabled robust, zero-touch coordination across multiple radio access technologies. The demonstrator scenario highlighted the viability of the approach in a realistic industrial setting, with \acp{sn} supporting mission-critical control with a latency between 2-3~ms and time-deterministic behavior, high-bandwidth \ac{iiot} sensors, and mobile logistics. Dynamic spectrum allocation allows the system to adapt to shifting demands, such as the arrival of a nomadic \ac{agv}, without compromising service quality. The next step will be to explore how the \ac{dsm} can be further extended topredict the spectrum needs of connected \acp{sn} based on trained datasets from an LLM. This can serve as a solution to reduce allocation conflicts in the \ac{dsm}.

\section*{Acknowledgment}
The authors acknowledge the financial support by the German \textit{Federal Ministry of Research, Technology and Space (BMFTR)} within the project Open6GHub \{16KISK004\ \& 16KISK010\}.

\bibliographystyle{ieeetr} % We choose the "plain" reference style
{%
\bibliography{references}
}%
\end{document}